\documentclass[reprint,amsmath,amssymb,aps,prb,]{revtex4-1}

\usepackage{graphicx}
\usepackage{amsmath}
\usepackage{dcolumn}
\usepackage{enumitem}
\usepackage{epstopdf}

\begin{document}

\title{Magnetization dynamics in proximity-coupled superconductor/ferromagnet/superconductor multilayers }

\author{I.~A.~Golovchanskiy$^{1,2,3}$, N.~N.~Abramov$^{2}$, V.~S.~Stolyarov$^{1,3}$, V.~I.~Chichkov$^{2}$, M.~Silayev$^{1,4}$, I.~V.~Shchetinin$^{2}$, A.~A.~Golubov$^{1,5}$, ,V.~V.~Ryazanov$^{2,6}$, A.~V.~Ustinov$^{2,7,8}$, M.~Yu.~Kupriyanov$^{1,9}$}

\affiliation{
$^1$ Moscow Institute of Physics and Technology, State University, 9 Institutskiy per., Dolgoprudny, Moscow Region, 141700, Russia; \\
$^2$ National University of Science and Technology MISIS, 4 Leninsky prosp., Moscow, 119049, Russia; \\
$^3$ Dukhov Research Institute of Automatics (VNIIA), 127055 Moscow, Russia; \\
$^4$ Department of Physics and Nanoscience Center, University of Jyv{\"a}skyl{\"a}, P.O. Box 35 (YFL), Jyv{\"a}skyl{\"a} FI-40014, Finland; \\
$^5$ Faculty of Science and Technology and MESA+ Institute for Nanotechnology, University of Twente, 7500 AE Enschede, The Netherlands; \\
$^6$ Institute of Solid State Physics (ISSP RAS), Chernogolovka, 142432, Moscow region, Russia; \\
$^7$ Physikalisches Institut, Karlsruhe Institute of Technology, 76131 Karlsruhe, Germany; \\
$^8$ Russian Quantum Center, Skolkovo, Moscow 143025, Russia; \\
$^9$ Skobeltsyn Institute of Nuclear Physics, MSU, Moscow, 119991, Russia
}%

\begin{abstract}
In this work, magnetization dynamics is studied in superconductor/ferromagnet/superconductor three-layered films in a wide frequency, field, and temperature ranges using the broad-band ferromagnetic resonance measurement technique.
It is shown that in presence of both superconducting layers and of superconducting proximity at both superconductor/ferromagnet interfaces a massive shift of the ferromagnetic resonance to higher frequencies emerges.
The phenomenon is robust and essentially long-range: it has been observed for a set of samples with the thickness of ferromagnetic layer in the range from tens up to hundreds of nanometers.
The resonance frequency shift is characterized by proximity-induced magnetic anisotropies: by the positive in-plane uniaxial anisotropy and by the drop of magnetization.
The shift and the corresponding uniaxial anisotropy grow with the thickness of the ferromagnetic layer.
For instance, the anisotropy reaches 0.27~T in experiment for a sample with 350~nm thick ferromagnetic layer, and about 0.4~T in predictions, which makes it a ferromagnetic film structure with the highest anisotropy and the highest natural resonance frequency ever reported. 
Various scenarios for the superconductivity-induced magnetic anisotropy are discussed. 
As a result, the origin of the phenomenon remains unclear.
Application of the proximity-induced anisotropies in superconducting magnonics is proposed as a way for manipulations with a spin-wave spectrum.
\end{abstract}

\maketitle

\section{Introduction}

Last two decades can be associated with a remarkable progress in areas of spin condensed matter physics, namely, in spintronics\cite{Zutic_RMP_76_323,Lu_IMR_61_456} and magnonics\cite{Kruglyak_JPDAP_43_264001,Lenk_PhysRep_507_107}.
Developments in spin physics have also advanced research in superconducting systems: by hybridizing superconducting and ferromagnetic orders intriguing physics emerges and new device functionality can be achieved, which is inaccessible in conventional systems.
Thus, superconducting spintronics\cite{Linder_NatPhys_11_307} can be viewed as a way for manipulation with spin states employing an interplay between ferromagnetic an superconducting spin orders. 
A long list of examples includes superconductor/ferromagnet/superconductor (S/F/S) josephson junctions\cite{Ryazanov_PRL_86_2427} that can be employed as phase pi-shifters\cite{Feofanov_NatPhys_6_539} and memory elements\cite{Vernik_IEEETAS_23_1701208,Golovchanskiy_PRB_94_214514}, F/S/F-based spin valves\cite{Lenk_PRB_96_184521}, and more complex long-range spin-triplet superconducting systems \cite{Robinson_Sci_329_59,Banerjee_NatComm_5_4771,Wang_NatPhys_6_389,Kapran_PRR_2_013167}.
Superconducting spintronics necessarily involves the superconducting proximity\cite{Buzdin_RMP_77_935} between ferromagnetic and superconducting subsystems.
On the other hand, superconducting magnonics can be viewed as manipulation with eigen-states of collective spin excitations via their interaction with a superconducting subsystem \cite{Dobrovolskiy_NatPhys_15_477,Golovchanskiy_AdvFunctMater_28_1802375,Golovchanskiy_AdvSci_6_1900435}.
In contrast to superconducting spintronics, in superconducting magnonics the proximity effect appears to be undesirable due to a possible suppression of fundamental characteristics of superconducting subsystem and consequently, degradation of the magnonic spectrum\cite{Golovchanskiy_JAP_127_093903}.

Recently, a qualitatively new manifestation of superconductor/ferromagnet hybridization has been reported, which in a way merges both areas the superconducting spintronics and the superconducting magnonics.
In Refs.~\cite{Li_ChPL_35_077401,Jeon_PRAppl_11_014061} a drastic increase
of the ferromagnetic resonance frequency has been observed in superconductor/ferromagnet/superconductor three-layers in presence of superconducting proximity between superconducting and ferromagnetic layers.
The origin of the phenomenon remains unclear. 
Possible explanations that has been proposed so far are attributed to incorporation of the spin-triplet superconducting pairing mechanism \cite{Li_ChPL_35_077401} or to an interplay of magnetization dynamics with the vortex/Meissner state of superconducting layers\cite{Jeon_PRAppl_11_014061}.
No convincing explanation has been provided so far.

In this paper, we report a detailed experimental study of the effect of superconducting proximity in S/F/S heterostructures on magnetization dynamics in the F-layer.
Experiments are performed using a broad-band ferromagnetic resonance (FMR) measurement technique in magnetic field, frequency and temperature domains.
This work is organized as follows.
Section II gives experimental details.
Section III provides experimental results: microwave ferromagnetic resonance absorption spectra at field-frequency domain at different temperatures and their quantitative analysis.
For a complete picture, we also suggest to review previous research studies on similar systems (see Refs.~\cite{Li_ChPL_35_077401,Jeon_PRAppl_11_014061,Jeon_PRB_99_144503}).
Section IV is devoted to discussion of experimental results where we state that the effect of superconducting proximity in S/F/S systems can not be explained employing concepts of the superconducting Meissner screening or of the vortex phase.
While the origin of the phenomena remains unclear at this stage, the authors suspect a contribution of spin-triplet superconductivity.
Section V demonstrates capabilities of the effect for manipulation of the spin-wave spectrum in S/F/S-based continuous films and magnonic crystals.

\section{Experimental details}

%
\begin{figure}[!ht]
\begin{center}
\includegraphics[width=1\columnwidth]{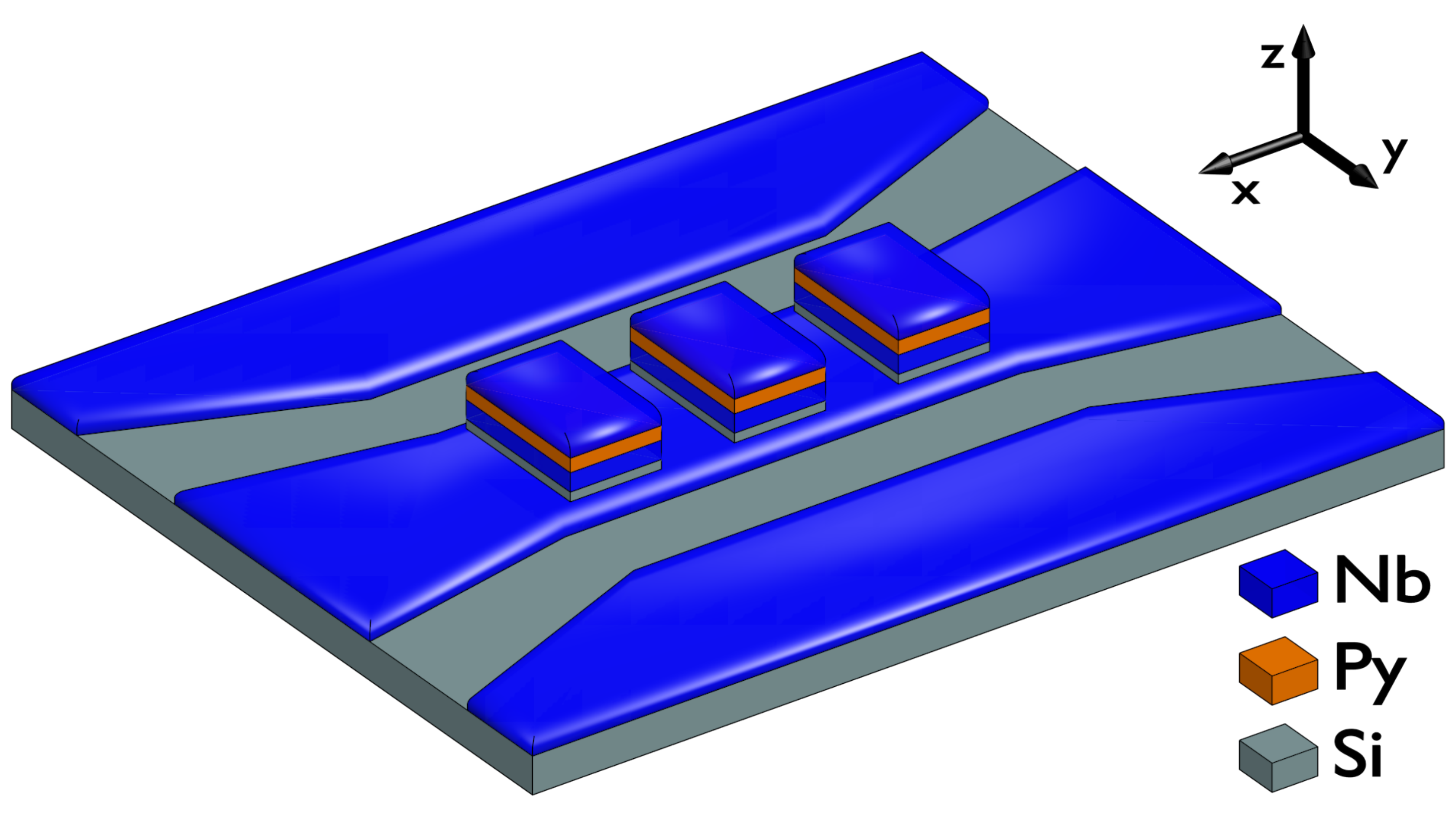}
\caption{
Schematic illustration of the investigated chip-sample.  
A series of S/F/S film rectangles is placed directly on top of the central transmission line of the co-planar waveguide.
Magnetic field $H$ is applied in-plane along the $x$-axis. 
}
\label{sam}
\end{center}
\end{figure}

Magnetization dynamics is studied by measuring the ferromagnetic resonance absorption spectrum using the VNA-FMR approach \cite{Neudecker_JMMM_307_148,Kalarickal_JAP_99_093909,Chen_JAP_101_09C104}.
A schematic illustration of the investigated chip-sample is shown in Fig.~\ref{sam}.
The chip consists of 150~nm thick superconducting niobium (Nb) co-planar waveguide with 50~Ohm impedance and 82-150-82~$\mu$m center-gap-center dimensions.
The waveguide is fabricated on top of Si/SiO$_x$ substrate using magnetron sputtering of Nb, optical lithography and plasma-chemical etching techniques.
A series of niobium/permalloy(Py=Fe$_{20}$Ni$_{80}$)/niobium (Nb/Py/Nb) film structures with lateral dimensions $X\times Y=50\times140$~$\mu$m and spacing of 25~$\mu$m along the $x-$axis is placed directly on top of the central transmission line of the waveguide using optical lithography, magnetron sputtering and the lift-off technique.
Importantly, deposition of Nb/Py/Nb three-layers is performed in a single vacuum cycle ensuring an electron-transparent metallic Nb/Py interfaces. 
A 20-nm-thick Si spacing is deposited between Nb co-planar and Nb/Py/Nb threelayers in order to ensure electrical insulation of the studied samples from the waveguide.
Five different samples has been fabricated and measured with different thickness of superconducting (S) and ferromagnetic (F) layers (see Tab.~\ref{Tab}).
One of samples was fabricated with an additional insulating (I) layer at one of S/F interfaces.

\begin{table}[h]
\begin{center}
\begin{tabular}{|c|c|c|c|c|}
\hline
Sample ID  & S(Nb) & F(Py) & I(AlO$_x$) & S(Nb) \\
\hline
S1 & 110 & 19 & 0 & 110  \\
\hline
S2 & 110 & 19 & 0 & 7 \\
\hline
S3 & 85 & 22 & 10 & 115\\
\hline
S4 & 140 & 45 & 0 & 140\\
\hline
S5 & 110 & 350 & 0 & 110\\
\hline
\end{tabular}
\caption{Parameters of studied samples.}
\label{Tab}
\end{center}
\end{table}

The experimental chip was installed in a copper sample holder and wire bonded to PCB with SMP RF connectors. 
A thermometer and a heater were attached directly to the holder for precise temperature control.
The holder was placed in a superconducting solenoid inside a closed-cycle cryostat (Oxford Instruments Triton, base temperature 1.2 K).
The response of experimental samples was studied by analyzing the transmitted microwave signal $S_{21}(f,H)$ with the VNA Rohde \& Schwarz ZVB20.
For exclusion of parasitic box resonance modes from consideration, all measured spectra $S_{21}(f,H)$ have been first normalized with $S_{21}(f)$ at $\mu_0H=0.3$~T, and then differentiated numerically in respect to $H$.
The response of experimental samples was studied in the field range from -0.22 T to 0.22 T, in the frequency range from 0 up to 18 GHz, and in the temperature range from 1.7 to 11 K.

\section{Experimental results: ferromagnetic resonance in proximity-coupled S/F/S systems}

\begin{figure*}[!ht]
\begin{center}
\includegraphics[width=0.66\columnwidth]{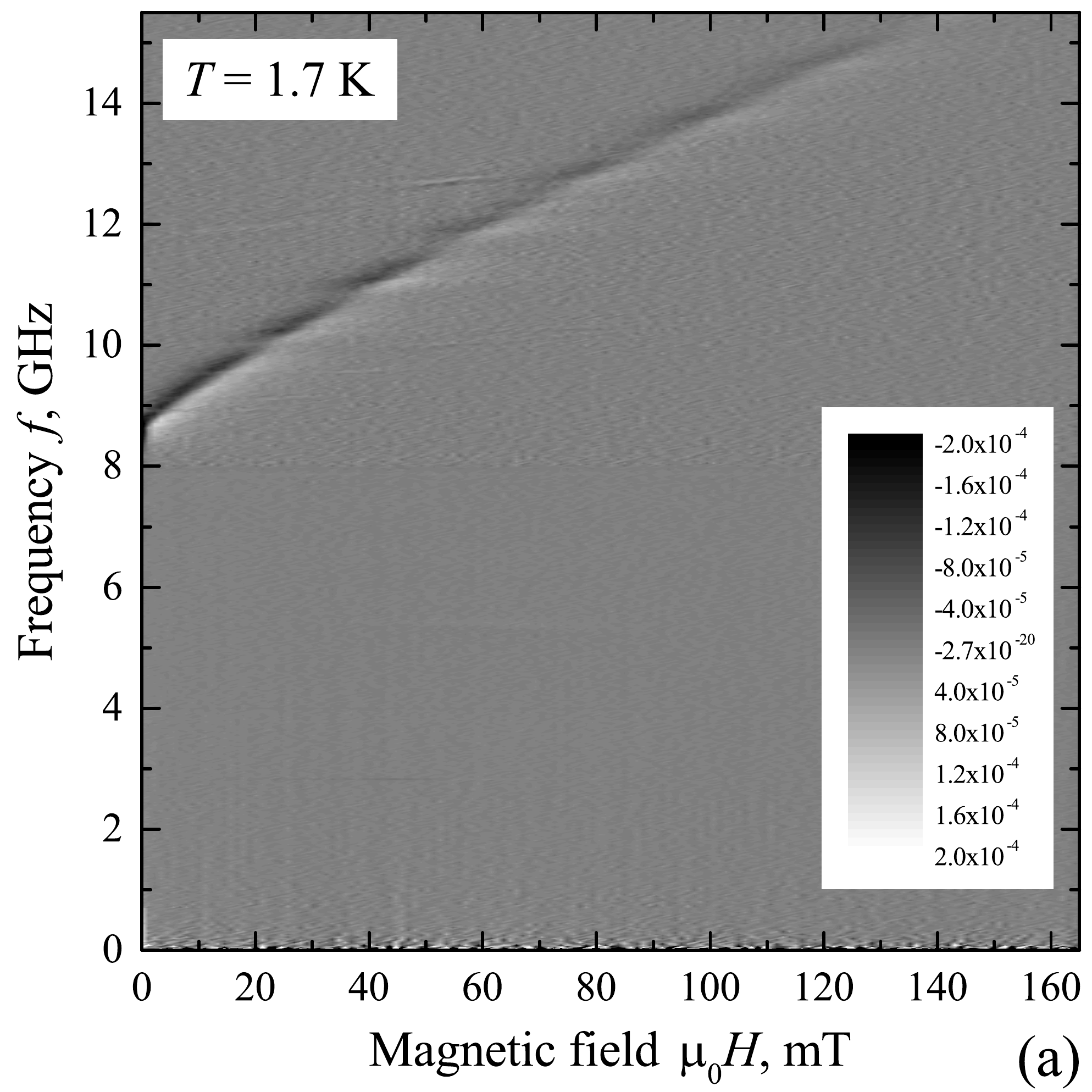}
\includegraphics[width=0.66\columnwidth]{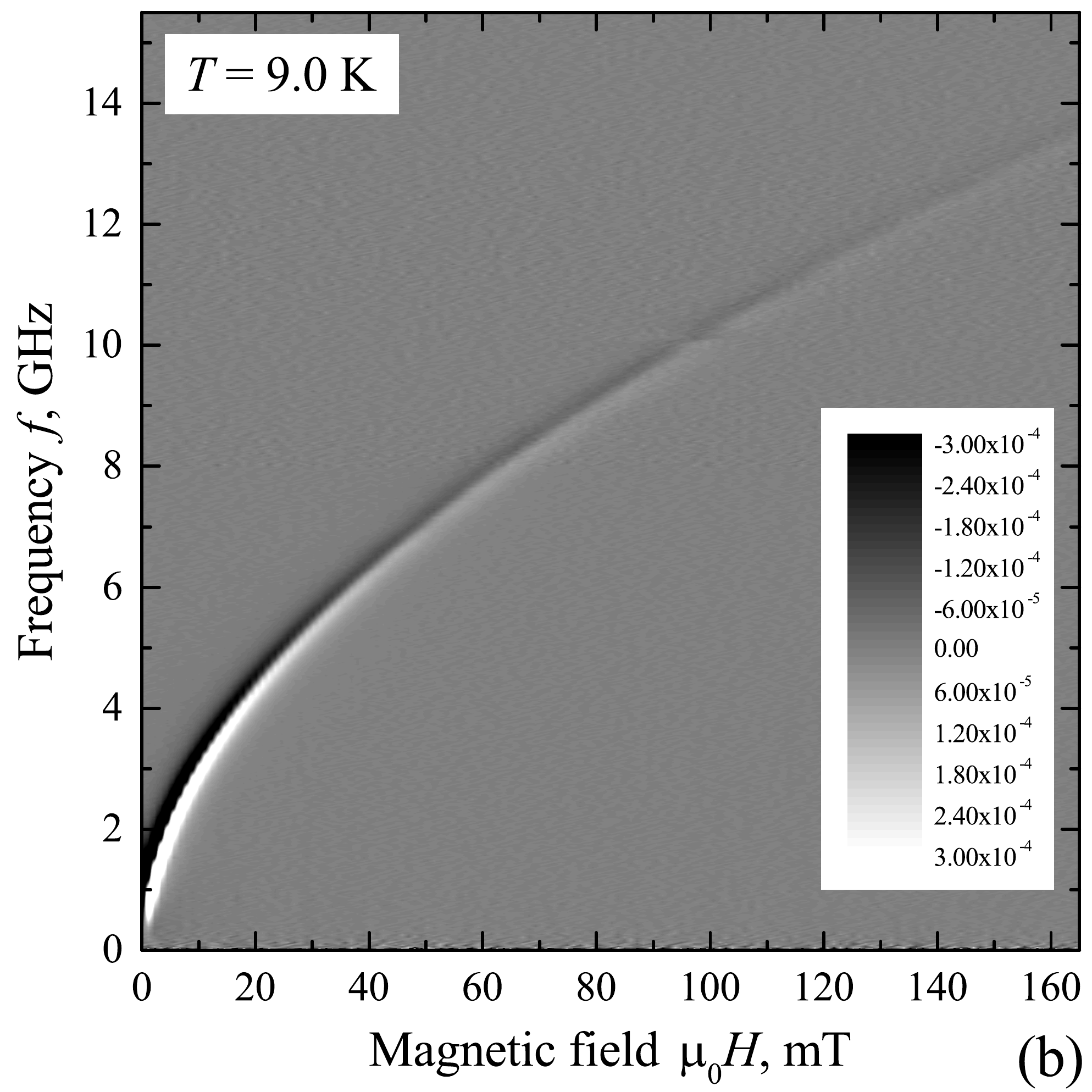}
\includegraphics[width=0.66\columnwidth]{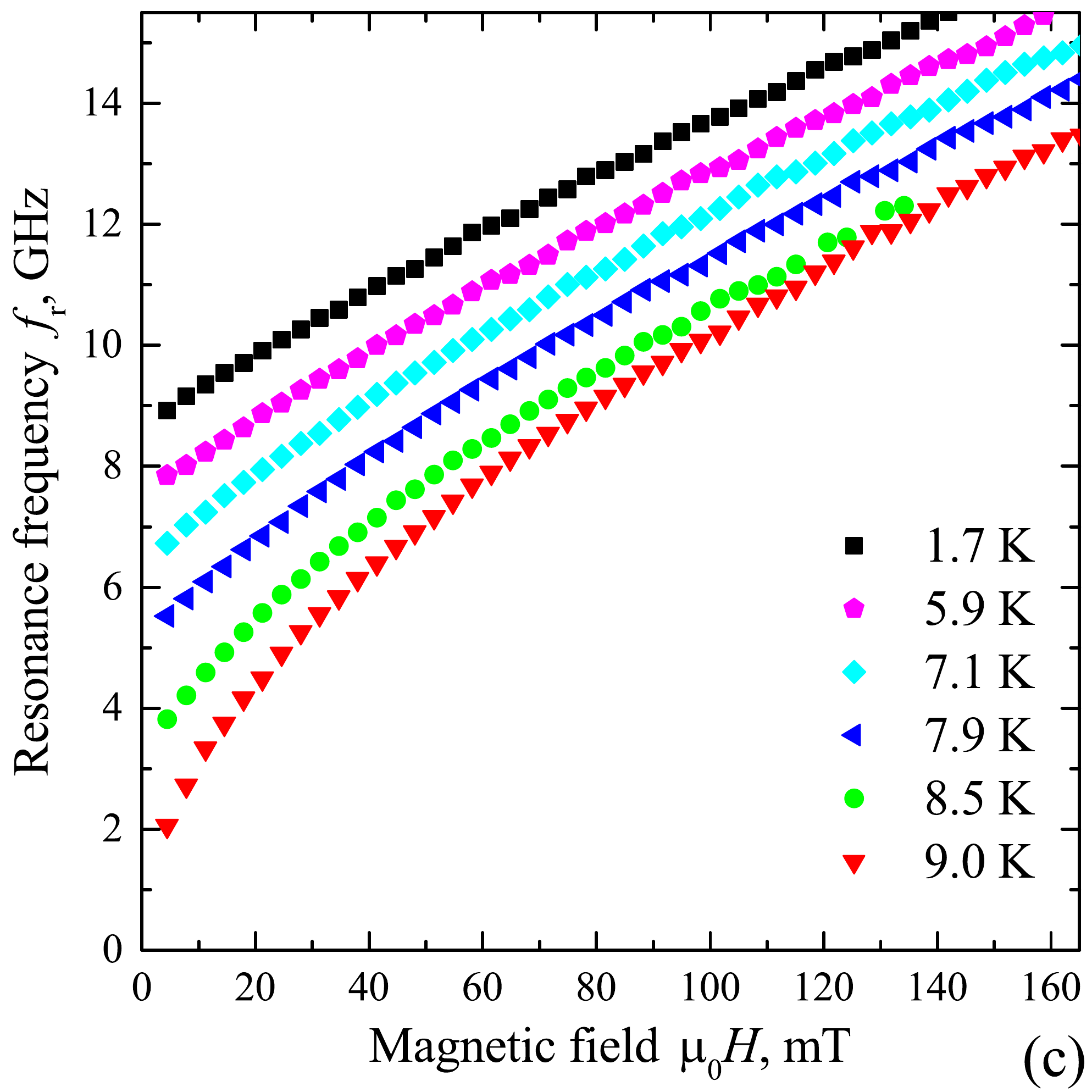}
\caption{
a,b) FMR absorption spectra $dS_{21}(f,H)/dH$ for S1 sample measured at $T=2$~K$>T_c$ (a) and $T=9$~K$\lesssim T_c$ (b).
The grayscale is coded in absolute units.
c) Dependencies of the FMR frequency on magnetic field $f_r(H)$ at different temperatures for S1 sample.
}
\label{Exp1}
\end{center}
\end{figure*}
\begin{figure*}[!ht]
\begin{center}
\includegraphics[width=0.66\columnwidth]{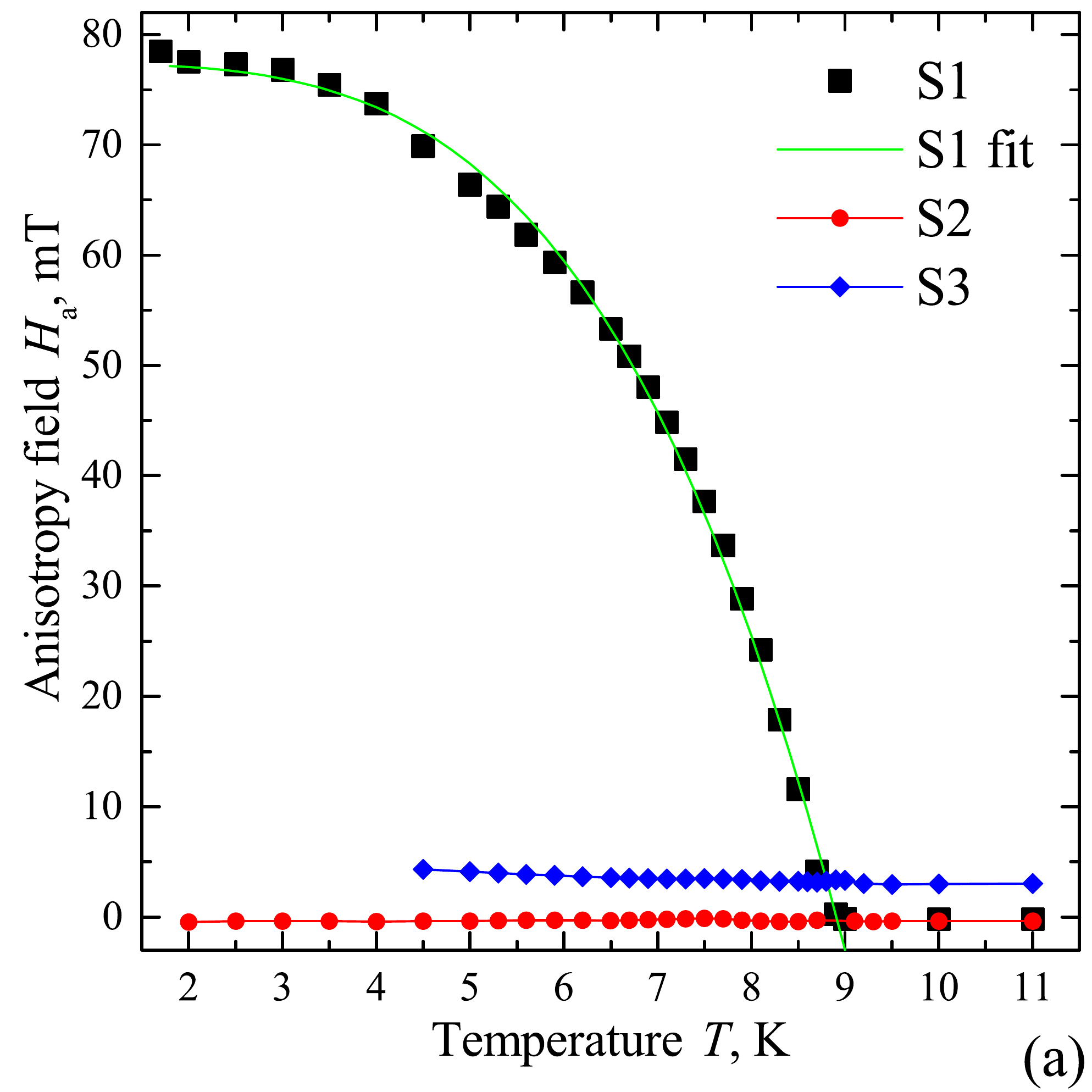}
\includegraphics[width=0.66\columnwidth]{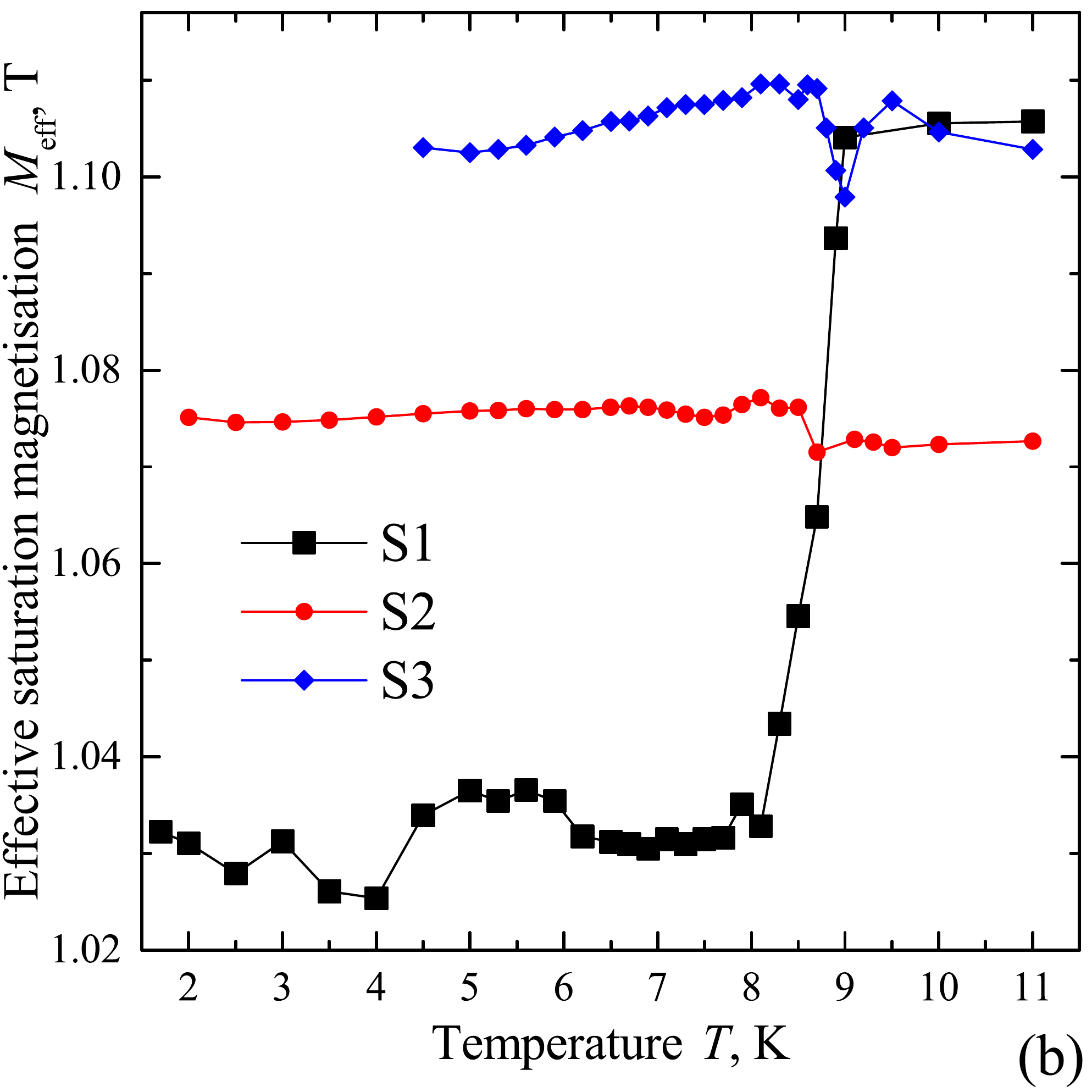}
\caption{The dependence of the anisotropy field $H_a$ (a) and effective magnetization $M_{eff}$ (b) on temperature.
Black square dots correspond to S1 S/F/S sample, red circular dots correspond to S2 S/F/s' sample, and blue diamond dots correspond to S3 S/F/I/S sample.
Green curve in (a) is the fit of $H_a(T)$ with Eq.~\ref{Ha}, which yields the following parameters: $\mu_0 H_{a0}=77$~mT, $T_c=9.0$~K, $p=3.7$.
}
\label{Fit1}
\end{center}
\end{figure*}

Figure~\ref{Exp1} illustrates the studied phenomenon using S(Nb)/F(Py)/S(Nb) sample with 110~nm thick Nb layers and 19~nm thick Py layer. 
This sample is referred to as S1.
Thickness of Py layer is selected for direct comparison of obtained results with previous research studies \cite{Li_ChPL_35_077401,Jeon_PRAppl_11_014061}.
Figures~\ref{Exp1}a,b show FMR absorption spectra $dS_{21}(f,H)/dH$ at $T=2$~K (a), which is far below the superconducting critical temperature $T_c$ of Nb, and at $T=9$~K (b), which corresponds to $T_c$.
Both spectra contain a single field-dependent spectral line, i.e., the FMR absorption line.
%
FMR absorption spectra at different temperatures have been fitted with the Lorentz curve and the dependencies of the resonance frequency on magnetic field $f_r(H)$ have been extracted.
Figure~\ref{Exp1}c collects resonance curves $f_r(H)$ that are measured at different temperatures.
Basically, Fig.~\ref{Exp1} demonstrates the essence of the phenomenon: it shows that upon decreasing the temperature below $T_c$ the resonance curve $f_r(H)$ shifts gradually to higher frequencies.
For instance, upon decreasing the temperature the frequency of the natural FMR $f_r(H=0)$ increases from about 0.5~GHz at $T\geq 9$~K to about 8.5~GHz at $T=1.7$~K.
 
FMR curves $f_r(H)$ in Fig.~\ref{Exp1}c follow the typical Kittel dependence for thin in-plane-magnetized ferromagnetic films at in-plane magnetic field: 
\begin{equation}
\left(2\pi f_r/\mu_0\gamma\right)^2=\left(H+H_a\right)\left(H+H_a+M_{eff}\right)
\label{Kit}
\end{equation}
where $\mu_0$ is the vacuum permeability, $\gamma=1.856\times10^{11}$~Hz/T is the gyromagnetic ratio for permalloy, $H_a$ is the uniaxial anisotropy field that is aligned with the external field, and $M_{eff}=M_s+M_a$ is the effective saturation magnetization, which includes the saturation magnetization $M_s$ and the out-of-plane anisotropy field $M_a$.
The fit of FMR curves in Fig.~\ref{Exp1}c with Eq.~\ref{Kit} yields the dependence of superconducting proximity-induced anisotropy fields $H_a$ and $M_{eff}$ on temperature given in Fig.~\ref{Fit1} with black squares.

Figure~\ref{Fit1} shows that at $T>T_c$ the anisotropy field is negligible $\mu_0 H_a\sim -2\times10^{-4}$~T and the effective magnetization is $\mu_0 M_{eff}\approx 1.1$~T.
These parameters are typical for permalloy thin films.
Also, at $T>T_c$ no dependence of $H_a$ and $M_{eff}$ on temperature is observed.
At $T<T_c$ upon cooling the anisotropy field $H_a$ increases gradually and reaches $\mu_0 H_a\approx 78$~mT at $T=2$~K.
This value is well consistent with previous studies on samples with the same thickness of Py layer \cite{Li_ChPL_35_077401,Jeon_PRAppl_11_014061}.
The dependence $H_a(T)$ can be characterized by fitting it with the following expression 
\begin{equation}
H_a=H_{a0}\left(1-(T/T_c)^p\right)
\label{Ha}
\end{equation}
where $H_{a0}$ is the effective anisotropy field at zero temperature, $T_c$ is the critical temperature, and $p$ is a free exponent parameter. 
The fit of $H_a(T)$ with Eq.~\ref{Ha} is shown in Fig.~\ref{Fit1}a with blue curve and yields the zero-temperature anisotropy $\mu_0 H_{a0}=77$~mT.

Importantly, the effective magnetization also demonstrates a temperature dependence: upon cooling $\mu_0 M_{eff}$ drops by about 70~mT.
Such effect was has not been obtained in previous studies\cite{Li_ChPL_35_077401,Jeon_PRAppl_11_014061} due to instrumental limitations.
The drop $-\Delta M_{eff}$ and the uniaxial anisotropy field $H_a$ at 2~K are roughly equal.
Thus, we state that superconductivity in S/F/S structure affects the magnetization dynamics by inducing positive in-plane anisotropy and by the drop of effective magnetization.

As the next step, following Ref.~\cite{Li_ChPL_35_077401}, we confirm that both superconducting layers are required for development of the effect of superconducting proximity on magnetization dynamics, and that electrical conductivity, i.e. the proximity, also is required to take place at both S/F interfaces.
The following S(Nb)/F(Py)/s'(Nb') sample is studied with 110~nm thick S(Nb) layer, 19~nm thick Py layer, which are similar to S1 sample, and thin 7~nm thick s'(Nb') layer .
This sample is referred to as S2 (see Tab.~\ref{Tab}). 
The upper s'(Nb') layer of S2 sample is argued to be non-superconducting due to its small thickness, below the superconducting coherence length and the London penetration depth, and due to the action of the inverse proximity effect.
Yet, the upper layer is expected to reproduce the microstructure of the upper Nb/Py interface. 
Basically, S2 sample represent S1 S/F/S sample with a removed superconducting layer.
FMR absorption spectra of S2 sample show no noticeable temperature dependence, which is consistent with previous studies \cite{Li_ChPL_35_077401}, and practically match with the spectrum of S1 sample at $T\gtrsim T_c$ (Fig.~\ref{Exp1}b).
Fitting procedures of FMR spectra and of resonance curves for S2 sample yield $H_a(T)$ and $M_{eff}(T)$ dependencies that a shown in Fig.~\ref{Fit1} with red circular dots.
The anisotropy field $H_a(T)$ in Fig.~\ref{Fit1}a is negligible, it varies in the range from $-5\times10^{-4}$~T to $-3\times10^{-4}$~T and shows no dependence on temperature.
The effective magnetization curve $M_{eff}(T)$, being at $\mu_0 M_{eff}\approx 1.072$~T at $T>T_c$, shows a minor increase by $\mu_0 \Delta M_{eff}\approx 3$~mT upon decreasing temperature and crossing $T_c$. 
Note that variation of $M_{eff}$ with temperature for S2 sample is opposite to one for S1 sample.

Next, the following S(Nb)/F(Py)/I(AlO$_x$)/S(Nb) sample is studied with thicknesses of Nb and Py layers similar to S1 and S2 samples, and additional insulating layer at one of S/F interfaces.
The sample is refereed to S3 (see Tab.~\ref{Tab}).
Basically, S3 sample represent S1 S/F/S sample with suppressed conductivity at one of S/F interfaces.
FMR absorption spectra of S3 sample shows no noticeable temperature dependence, which is consistent with previous studies \cite{Li_ChPL_35_077401}.
Blue diamond dots in Fig.~\ref{Fit1} show $H_a(T)$ and $M_{eff}(T)$ dependencies for S3 sample.
The anisotropy field $H_a(T)$ in Fig.~\ref{Fit1}a is negligible, though is slightly higher than one for S1 and S2 samples.
It varies in the range from 3 to 5~mT and shows insignificant dependence on temperature.
The effective magnetization curve $M_{eff}(T)$, varies in the range from 1.1 up to 1.2~T and shows a minor drop by $\mu_0 \Delta M_{eff}\approx 10$~mT in vicinity to $T_c$.
Therefore, with S2 and S3 samples we confirm that both superconducting layers are required for development of the effect of superconducting proximity on magnetization dynamics and that superconducting proximity is required to take place at both S/F interfaces.

\begin{figure*}[!ht]
\begin{center}
\includegraphics[width=0.66\columnwidth]{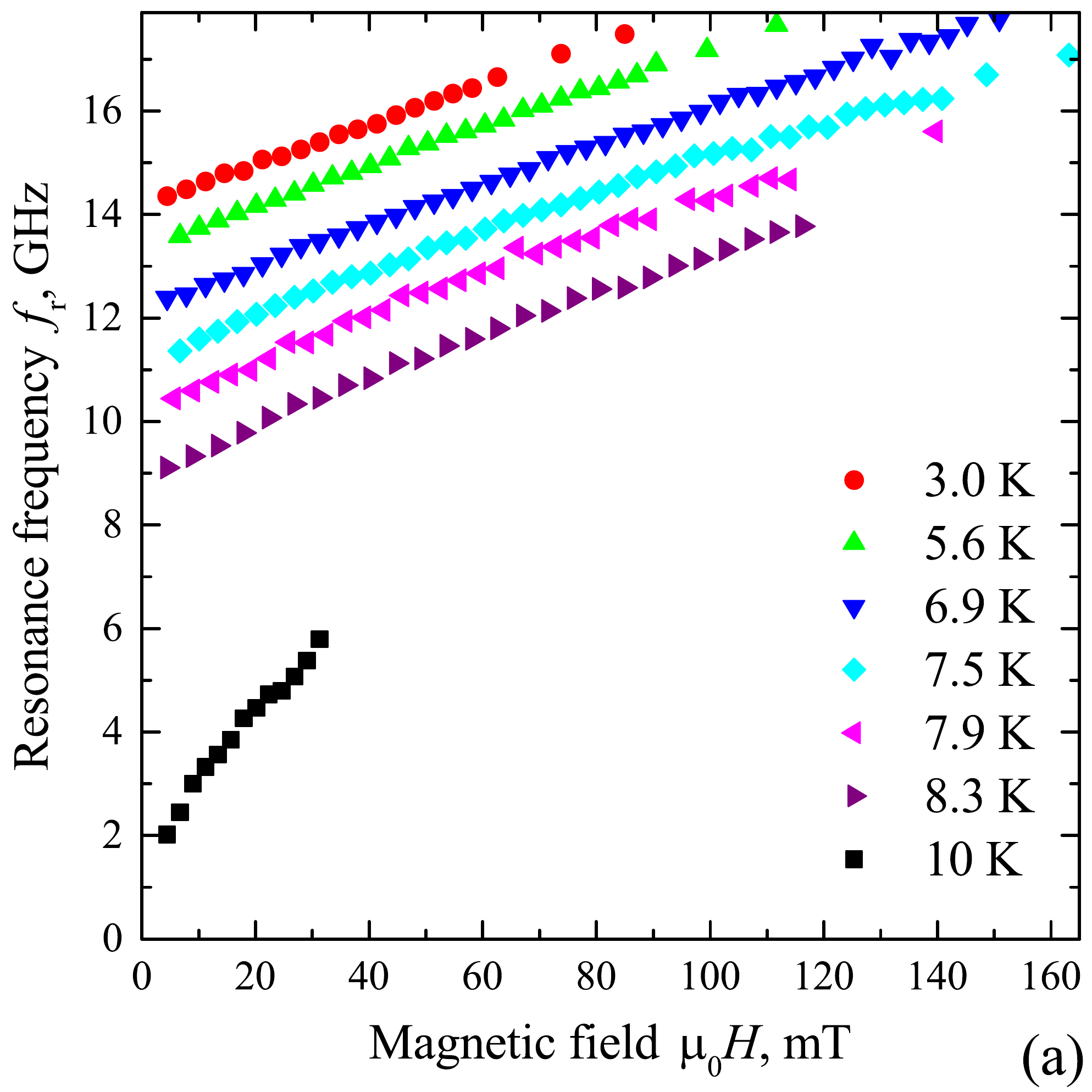}
\includegraphics[width=0.66\columnwidth]{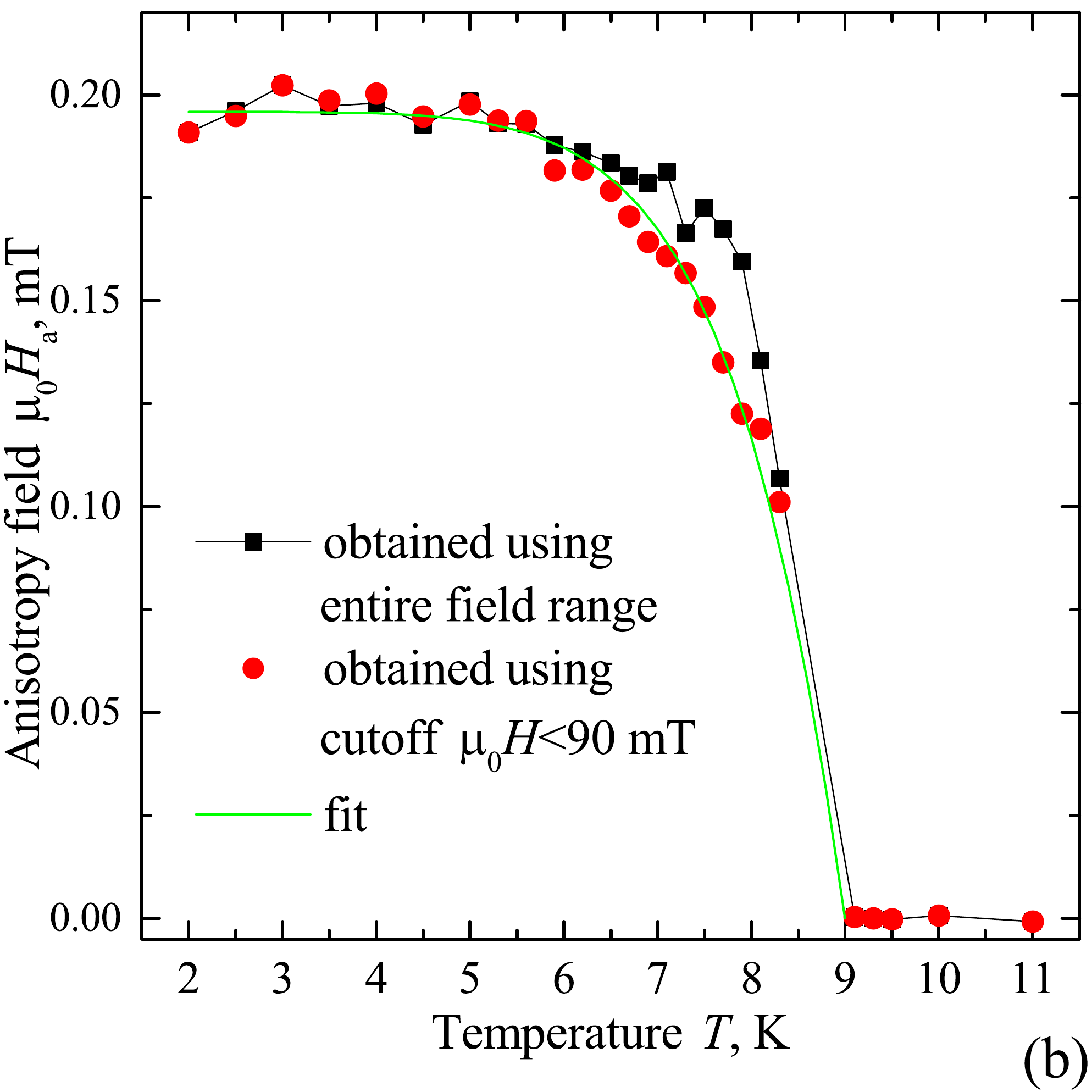}
\includegraphics[width=0.66\columnwidth]{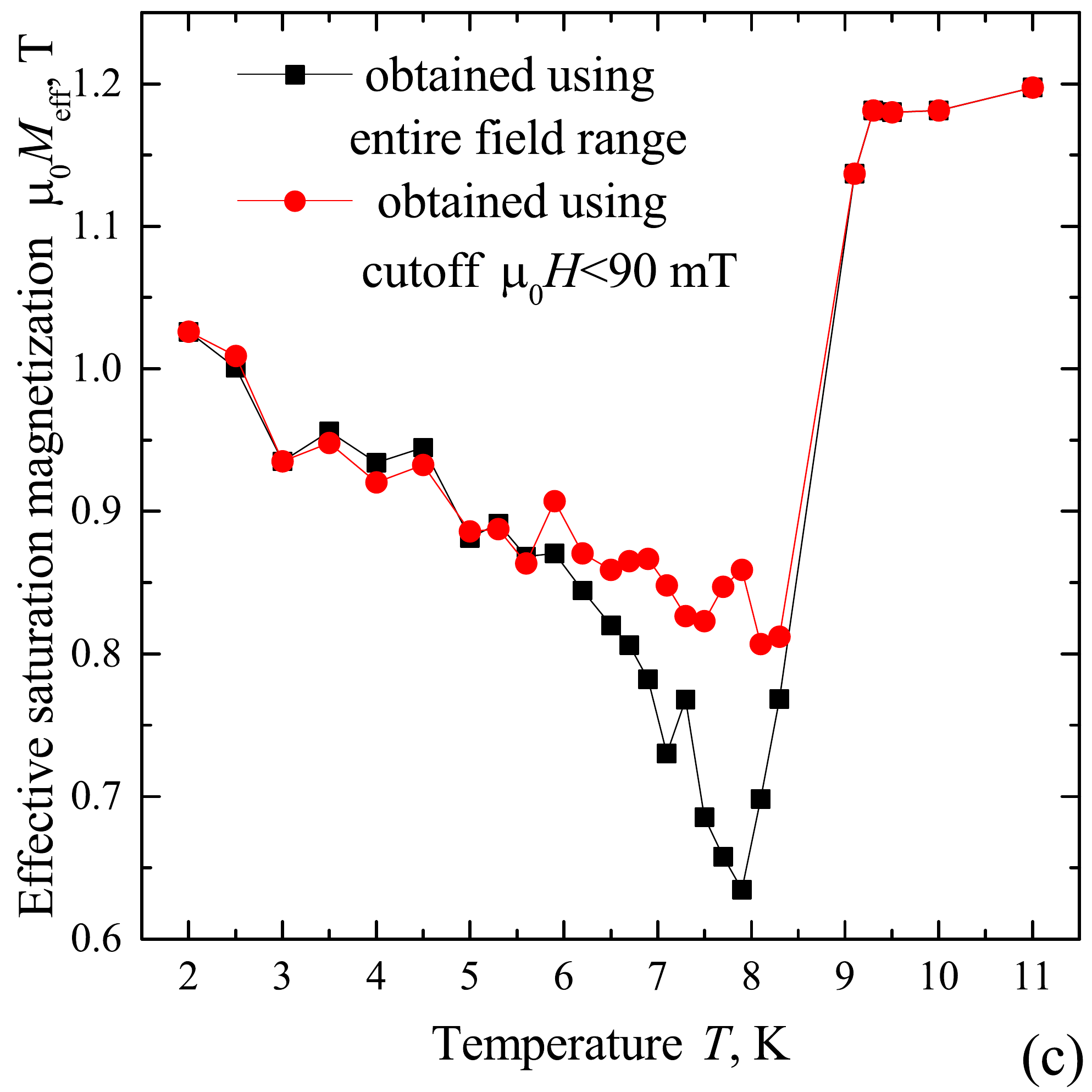}
\caption{
a) Dependencies of the FMR frequency on magnetic field $f_r(H)$ at different temperatures for S4 sample.
b,c) The dependence of the anisotropy field $H_a$ (b) and effective magnetization (c) on temperature.
The data in b,c that is shown with black square dots was obtained by fitting $f_r(H)$ in the entire field range from 0 up to 200 mT.
The data in b,c that is shown with red circular dots was obtained by fitting $f_r(H)$ in the cut-off field range from 0 up to 90 mT.
Green curve in (b) shows the fit of $H_a(T)$, which is obtained using the cut-off field range, with Eq.~\ref{Ha}, which yields the following parameters: $\mu_0 H_{a0}=196$~mT, $T_c=9.0$~K, $p=7.7$.
}
\label{Exp3}
\end{center}
\end{figure*}

As a crucial step, the dependence of phenomenon on the thickness of the F-layer is revealed. 
Figure~\ref{Exp3} demonstrates this dependence with a different S(Nb)/F(Py)/S(Nb) sample with 140~nm thick Nb layers and 45~nm thick Py layer. 
This sample is referred to as S4 (see Tab.~\ref{Tab}).
Figure~\ref{Exp3}a collects resonance curves $f_r(H)$ that are measured at different temperatures.
It shows that upon decreasing the temperature below $T_c$ the resonance curve $f_r(H)$ shifts gradually to higher frequencies following the same trend as for S1 sample.
Comparison of Fig.~\ref{Exp3}a with Fig.~\ref{Exp1}a immediately indicates that the effect of the superconducting proximity in S/F/S systems on magnetization dynamics is substantially stronger for the thicker S4 sample: upon decreasing the temperature the frequency of the natural FMR increases from about 1~GHz at $T=10$~K up to about 14.5~GHz at $T=3$~K.
In other terms, by increasing the thickness of the F layer by a factor of 2.3 the enhancement of the natural FMR frequency of S/F/S sample in superconducting state at $T\ll T_c$ has increased by a factor of 1.6.

The fit of FMR curves in Fig.~\ref{Exp3}a with Eq.~\ref{Kit} yields the dependence of superconducting proximity-induced anisotropy fields $H_a$ and $M_{eff}$ on temperature that are given in Fig.~\ref{Exp3}b,c with black squares.
Figure~\ref{Exp3}b shows that at $T>T_c$ the anisotropy field is negligible as in case of S1, S2 and S3 samples.
At $T<T_c$ upon cooling the anisotropy field $H_a$ increases gradually and reaches $\mu_0 H_a\approx 200$~mT at $T=2$~K.

The temperature dependence of the effective magnetization $M_{eff}(T)$ given in Figure~\ref{Exp3}c is more complex and is qualitatively different from one for S1 sample.
Upon cooling $\mu_0 M_{eff}$ first drops from 1.2~T at $T>T_c$ to about 0.6~T at $T\lesssim T_c$ and than increases gradually up to about 1.03~T at $T=2~K$.
We argue that such temperature dependence can be explained by field dependence of proximity-induced parameters.
Indeed, at fixed $T<T_c$ at upper-right section of a resonance absorption spectrum $S_{21}(f,H)$ superconductivity is partially suppressed by external field and microwave radiation, and therefore $H_a$ is expected to be reduced while $M_{eff}$ is expected to be increased as compared to lower-left section of the spectrum.
This phenomenon can be illustrated by fitting of FMR curves in Fig.~\ref{Exp3}a with Eq.~\ref{Kit} in the limited field range.
Red circular dots in Fig.~\ref{Exp3}b,c show temperature dependencies of $H_a$ and $M_{eff}$ obtained by fitting only part of FMR curves at $\mu_0 H<90$~mT.
Figure~\ref{Exp3}c shows that the drop of $M_{eff}$ at $T\lesssim T_c$ is significantly reduced: upon cooling $\mu_0 M_{eff}$ first drops from 1.2~T at $T>T_c$ to about 0.8~T at $T\lesssim T_c$ and than increases gradually up to about 1.03~T at $T=2~K$.
Green curve in Figure~\ref{Exp3}b shows the fit of $H_a(T)$, which is obtained using the cut-off field range, with Eq.~\ref{Ha}. 
The fit yields the zero-temperature anisotropy $\mu_0 H_{a0}=196$~mT.
Overall, the drop $-\Delta M_{eff}$ and the induced $H_a$ at 2~K are are roughly equal as in case of S1 sample: the anisotropy field $\mu_0 H_{a0}=196$~mT while the drop of the effective magnetization $\mu_0 \Delta M_{eff}\approx -170$~mT.

Importantly, FMR parameters of the S1 sample, $H_a(T)$ and $M_{eff}(T)$ in Fig.~\ref{Fit1}, are mostly unchanged when obtained using the same limited range of magnetic fields $\mu_0 H<90$~mT.
This fact can be explained by frequency dependence of proximity-induced anisotropy fields. 
Indeed, resonance frequencies for S1 sample are typically by a factor of 2 lower than for S4 sample.
Therefore, the superconducting state of S-layers in S1 sample is less affected by microwave radiation than in S4 sample.

\begin{figure*}[!ht]
\begin{center}
\includegraphics[width=0.66\columnwidth]{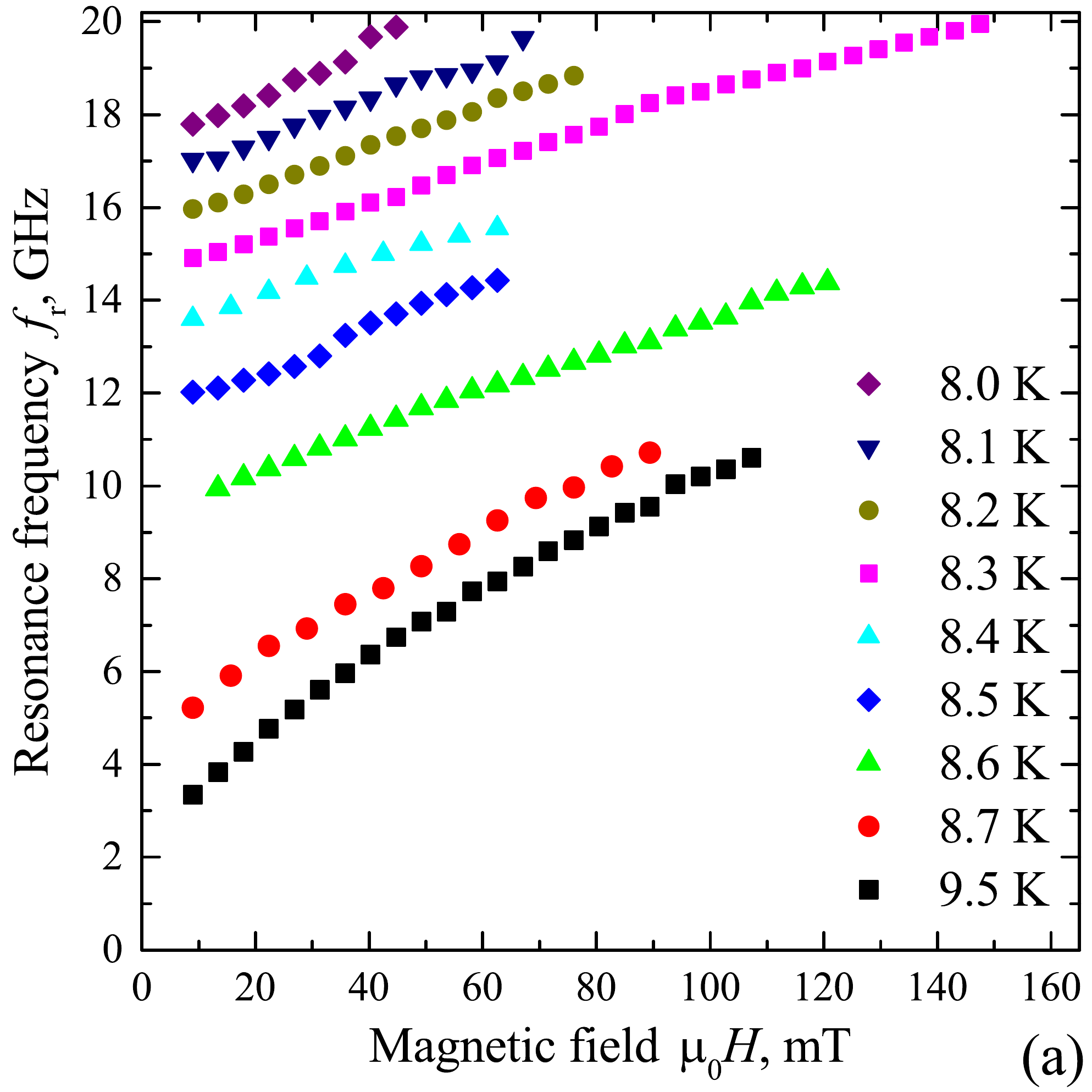}
\includegraphics[width=0.66\columnwidth]{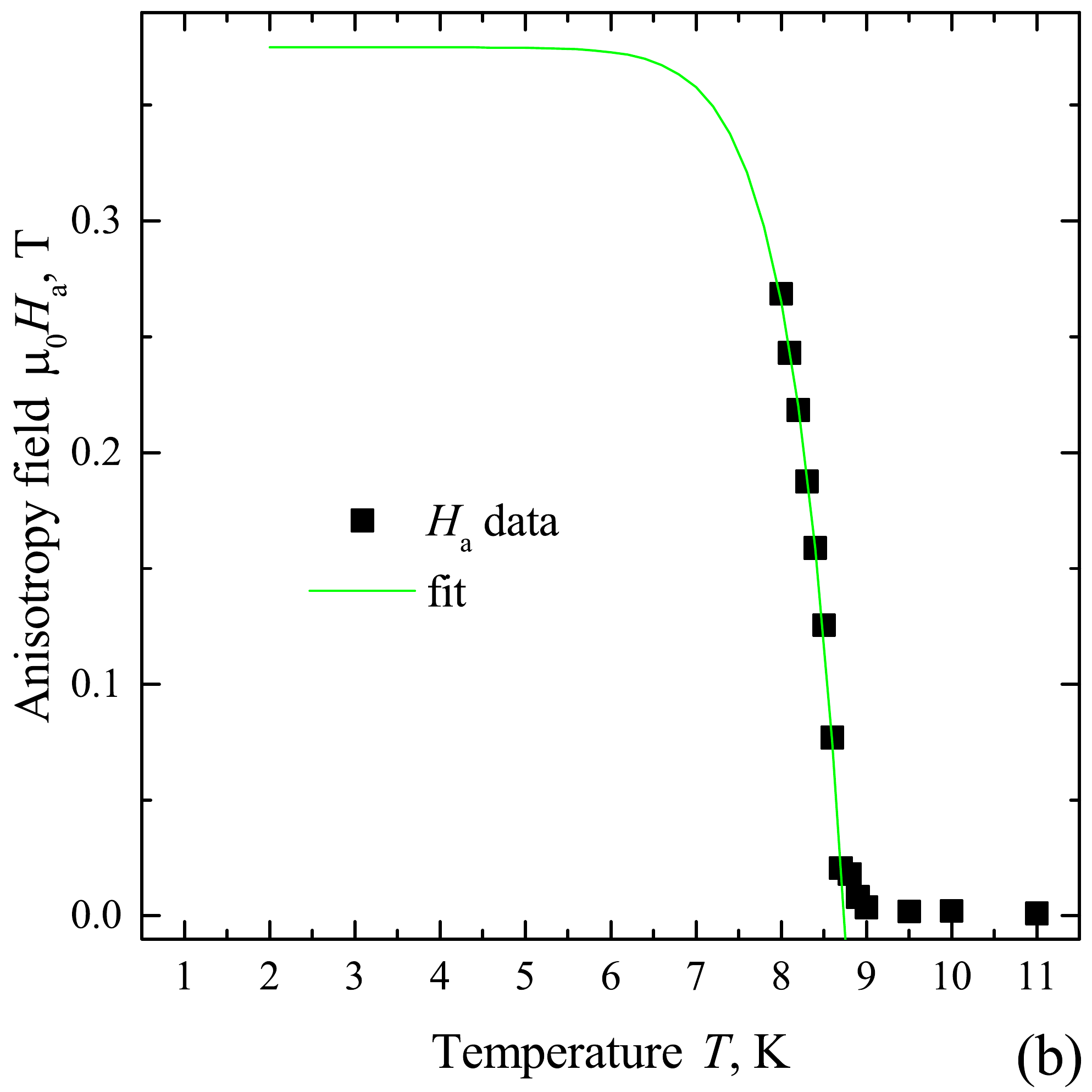}
\caption{
a) Dependencies of the FMR frequency on magnetic field $f_r(H)$ at different temperatures for S5 sample.
b) The dependence of the anisotropy field $H_a$ on temperature.
Green curve in (b) shows the fit of $H_a(T)$ with Eq.~\ref{Ha}, which yields the following parameters: $\mu_0 H_{a0}=375$~mT, $T_c=8.74$~K, $p=13.9$
}
\label{Exp6}
\end{center}
\end{figure*}

Figure~\ref{Exp6} demonstrates the effect of the superconducting proximity in S/F/S systems on magnetization dynamics for a different S(Nb)/F(Py)/S(Nb) sample with a radically thicker 350~nm thick Py layer. 
This sample is referred to as S5 (see Tab.~\ref{Tab}).
Figure~\ref{Exp6}a collects resonance curves $f_r(H)$ that are measured at different temperatures; it shows that upon decreasing the temperature below $T_c$ the resonance curve $f_r(H)$ shifts gradually to higher frequencies following the same trend as for S1 and S4 samples.
However, the enhancement of the FMR frequency upon decreasing temperature at $T<T_c$ is so intense that the FMR curve approaches the instrumental frequency band limit already at $T\sim8$~K (note the temperature range in legend of Fig.~\ref{Exp6}a).
Comparison of Fig.~\ref{Exp6}a with Figs.~\ref{Exp1}a~and~\ref{Exp3}a confirms that the effect of the superconducting proximity in S/F/S systems on magnetization dynamics enhances with growing thickness of the F-layer.
Upon decreasing the temperature the frequency of the natural FMR of S5 sample increases from about 1~GHz at $T>T_c$ up to about 17~GHz already at $T=8$~K.
Proximity to the superconducting critical temperature, insufficient signal-to-noise ratio, parasitic box modes, did not allow to fit resonance curves considering both $H_a$ and $M_{eff}$ in Eq.~\ref{Kit} as fitting parameters.
Therefore, the fitting routine was modified for S5 sample as follows.
First, $f_r(H)$ curves have been fitted at $T>T_c$ with Eq.~\ref{Kit}.
The fit yields $\mu_0 M_{eff}\approx 1.076$~T and $\mu_0 H_a \sim 1$~mT.
Next, $f_r(H)$ curves at $T<T_c$ have been fitted with Eq.~\ref{Kit} considering magnetization fixed at $\mu_0 M_{eff}=1.076$~T and considering $H_a$ as the only fitting parameter.
The dependence $H_a(T)$ is given in Fig.~\ref{Exp6}b with black squares.
It shows that the effective anisotropy field reaches $\mu_0 H_a\approx 0.27$~T at 8~K.
Note that by fixing $M_{eff}$ the so-obtained anisotropy field $H_a$ is expected to be underestimated since according to $M_{eff}(T)$ dependencies for S1 and S4 samples $M_{eff}$ should actually drop at $T<T_c$.
Green curve in Figure~\ref{Exp6}b shows the fit of $H_a(T)$ with Eq.~\ref{Ha}. 
The fit yields the extrapolated zero-temperature anisotropy $\mu_0 H_{a0}=375$~mT, which is also expected to be underestimated.

Summarizing experiential findings, superconductivity in S/F/S three-layers shifts the FMR to higher frequencies.
The shift can be quantified by the proximity-induced positive in-plane anisotropy $H_a$ and by a drop of effective magnetization $M_{eff}$.
Both $H_a$ and the drop of $M_{eff}$ are roughly equal and are field-, frequency- and temperature-dependent.
The phenomenon requires both superconducting layers of S/F/S and presence of superconducting proximity at both S/F interfaces. 
The phenomenon shows a dependence on the thickness of the F-layer: for thicker F-layer the shift of the FMR frequency is substantially stronger.
In addition, it should be noted that
(i) no dependence of the FMR spectrum on the input power has been observed in the range of input power from -15~dB to 0~dB;
(ii) all measured spectra for all samples are field-reversible; and
(iii) no dependence of the FMR linewidth on experimental parameters could be noted owing partially to insufficient signal-to-noise ratio.
As a final remark it should be noted that, technically, samples S4 and S5 demonstrate the highest natural FMR frequencies and corresponding in-plane anisotropies for in-plane magnetized ferromagnetic film systems ever reported (see, for instance, Ref.~\cite{Li_SciRep_5_17023} for comparison).

\section{Discussions: possible origin of proximity-induced anisotropies in S/F/S systems}

A natural initial guess for the origin of the effect of superconducting proximity in S/F/S systems on magnetization dynamics is the Meissner screening of external field, the so-called lensing effect \cite{Golovchanskiy_AdvSci_6_1900435,Schmidt}.
For instance, one could employ fluxometric or magnetometric demagnetizing factors\cite{Aharoni_JAP_87_6564,Golovchanskiy_JAP_123_173904} of the system for estimation of a hypothetical diamagnetic moment in Nb layers that induces magnetostatic field $H_a$.
However, this estimation is not required since the following set of unfulfilled conditions points towards irrelevance of the lensing effect in discussed experiments:
(i) In case of the lensing effect the induced $H_a$ is not a constant but a field-dependent quantity \cite{Golovchanskiy_AdvSci_6_1900435}.
(ii) In case of the lensing effect the induced $H_a$ should decrease with increasing thickness of the F-layer.
(iii) The lensing effect should hold for S/F/I/S structure (S3 sample)  and should be only halved for S/F structure (S2 sample) .
(iv) The field that is induced by the lensing effect can not exceed the first critical field, which in Nb is about 100~mT (see values of $H_a$ in Figs.~\ref{Exp3}b~and~\ref{Exp6}b).
None of the above hypothetical effects does take place.
In addition, consideration of the lensing effect does not clarify possible origin of the drop of magnetization $\Delta M_{eff}$ at $T<T_c$ in Figs.~\ref{Fit1}b~and~\ref{Exp3}c.

In fact, S/F (S2) and S/F/I/S (S3) structures may evidence the effect of Meissner screening on precessing magnetization in thin film geometry.
Meissner screening is expected to show itself in the absence of the in-plane anisotropy and in the presence of small negative out-of-plane uniaxial anisotropy.
The later might be indicated by a small variation of $M_{eff}$ at $T\sim T_c$ in Figs.~\ref{Fit1}c~and~\ref{Exp3}c.

The next hypothetical candidate for impact on magnetostatic state of the F-layer is the vortex phase.
The following set of unfulfilled conditions evidence that the vortex phase can not have any effect on magnetization dynamics:
(i) The effect of the vortex phase should hold for S/F/I/S structure (S3 sample).
(ii) Presence of the vortex phase that is induced by the external magnetic field should, in the first place, lead to hysteresis in the absorption spectrum due to pinning \cite{Golovchanskiy_PRAppl_11_044076}.
(ii) The density of a vortex phase that is induced by the external field is expected to be field-dependent leading to field-dependence of hypothetical vortex-phase-induced anisotropies.
(iii) Presence of the vortex phase in superconducting thin films induces only insignificant total magnetic moments and corresponding stray fields. 
In addition, low expected density and arbitrary nature of the out-of-plane vortex phase unfavor its possible contribution.

Mechanisms that are considered above are limited to magnetostatic interactions between F- and S-subsystems.
Alternative explanations imply electronic correlations between superconducting and ferromagnetic subsystems.
For instance, in Refs.~\cite{Mironov_APL_113_022601,Volkov_PRB_99_144506,Devizorova_PRB_99_104519} the electromagnetic proximity effect and spin polarization in planar superconductor-ferromagnet structures are discussed.
The electromagnetic proximity effect implies presence of the superconducting condensate in the ferromagnetic layer and induction of screening currents in the S/F system as a response on magnetic moment\cite{Stolyarov_ScAdv_4_eaat1061} rather than on magnetic field.
While in general the electromagnetic proximity effect is diamagnetic and induces magnetic field that counteracts the magnetization, at certain thicknesses of the F-layer the so-called paramagnetic electromagnetic proximity effect can take place, which induces magnetic field along the magnetization \cite{Mironov_APL_113_022601}.
However, large thickness of F-layers in our experiments of 20, 40 and 350~nm in comparison to the typical electron correlation length of singlet pairs in ferromagnets\cite{Eschrig_PhysTod_64_43,Eschrig_RPP_78_104501,Blamire_JPCM_26_453201} $\xi_F\sim 1$~nm, and predicted oscillating behaviour of the sign of induced field with the thickness of the F-layer rule-out contribution of the electromagnetic proximity effect on magnetization dynamics in considered S/F/S systems.

Also, one can rule-out possible contribution of the spin-inverse proximity effect or the so-called spin-screening \cite{Bergeret_EPL_66_111,Dahir_PRB_100_134513}.
The spin-screening considers accumulation of spins with polarization opposite to F-magnetization in a thin layer of the S-subsystem of the order of the coherence length in vicinity to the S/F interface. 
Such spin orientation could possibly produce stray fields of a required direction along magnetization in the F-layer.
Yet, owing to thin film geometry and small demagnetizing factors\cite{Aharoni_JAP_87_6564,Golovchanskiy_JAP_123_173904} of the system an implausibly large magnetization of the spin-polarized area is required for induction of the observed $H_a$, which is far above superconducting critical fields.

Another possible explanation for the effect of superconducting proximity in S/F/S systems on magnetization dynamics is provided in the very first report of the effect.
In Ref.~\cite{Li_ChPL_35_077401} it is proposed that the effective anisotropy field is produced due to interaction of magnetization with spin-polarized spin-triplet superconducting electrons via the spin-transfer torque mechanism\cite{Brataas_NatMat_11_372,Brataas_arXiv,Ralph_JMMM_320_1190,Sankey_PRL_96_227601}.
This mechanism requires presence of spin-triplet superconducting pairs as a necessary ingredient.
In Ref.~\cite{Li_ChPL_35_077401} it is proposed that the spin-triplet superconductivity is induced by the dynamically precessing magnetization in accordance with the Ref.~\cite{Houzet_PRL_101_057009}.
However, such mechanism requires large frequency of magnetization precession that should be comparable to the depairing frequency and is inconsistent with the frequency range of reported results. 

Thus, we state that at this stage even a qualitative explanation of the effect of superconducting proximity in S/F/S systems on magnetization dynamics is unavailable.
Yet, long-range nature of the phenomenon and the mandatory S/F/S symmetry of the phenomenon are signatures for a role of spin-triplet superconductivity\cite{Eschrig_PhysTod_64_43}.

\section{Prospects of the proximity effect for application in magnonics}

\begin{figure*}[!ht]
\begin{center}
\includegraphics[width=0.66\columnwidth]{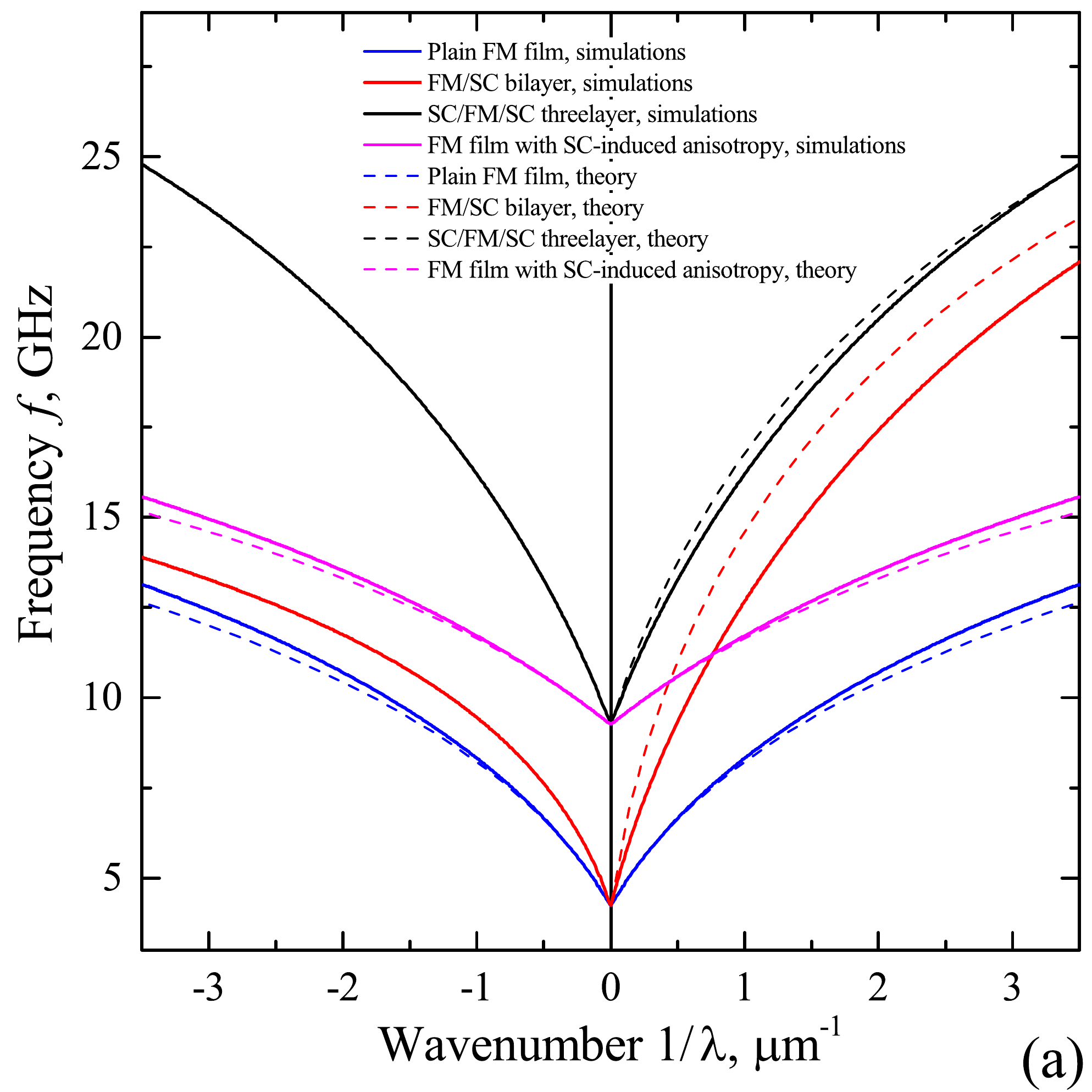}
\includegraphics[width=0.66\columnwidth]{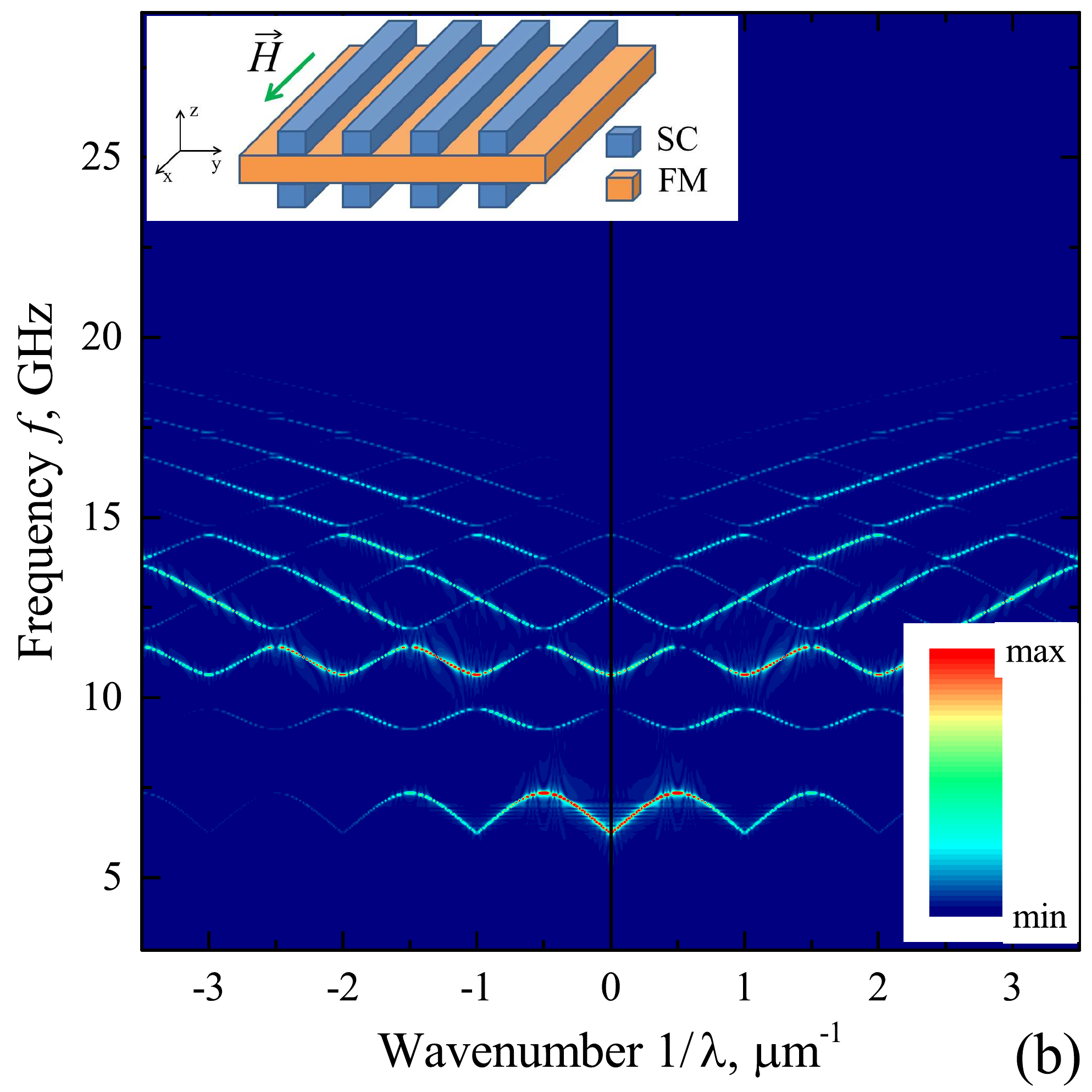}
\includegraphics[width=0.66\columnwidth]{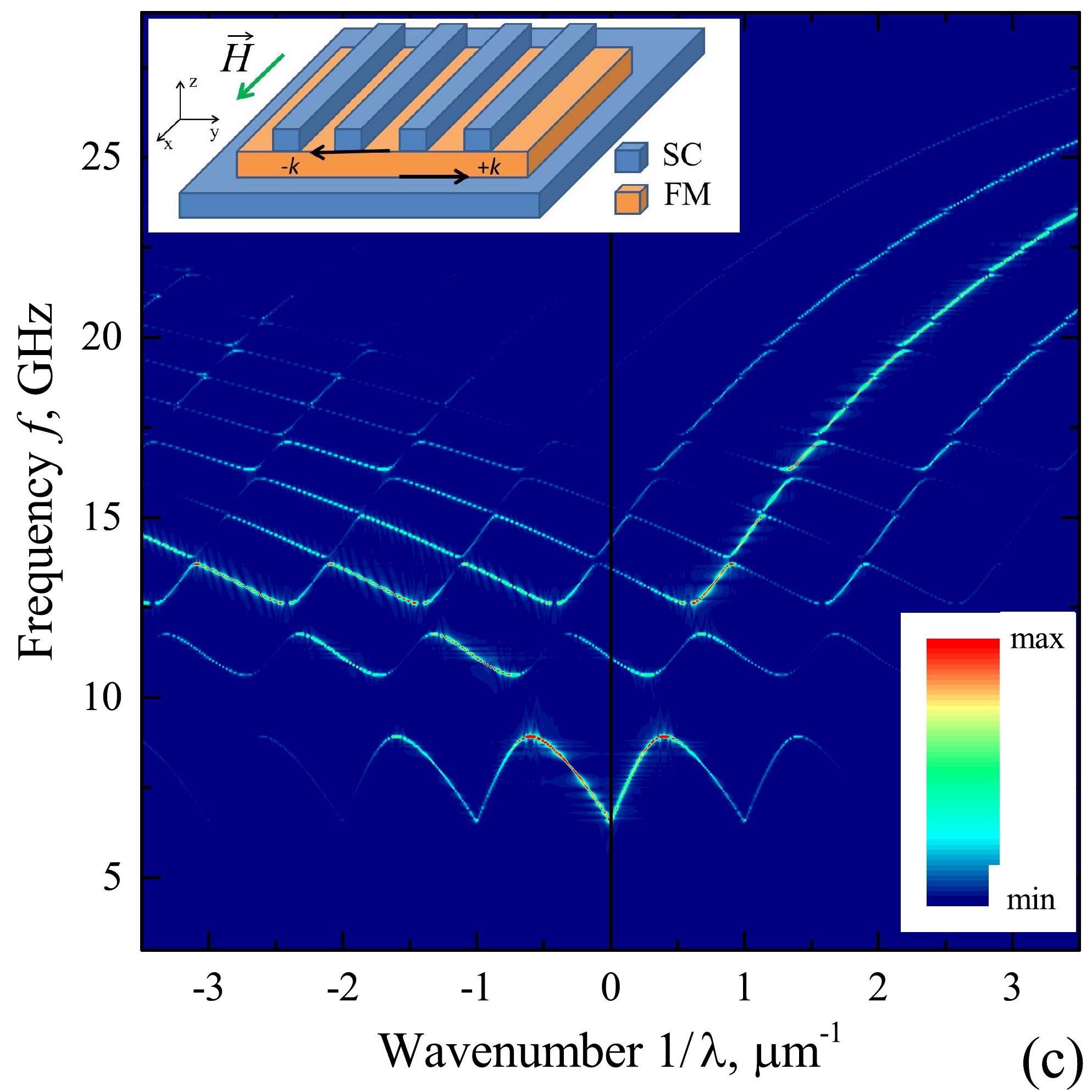}
\caption{a) Dispersion curves for MSSW that propagate in continuous F, S/F, and S/F/S films.
Dispersion curves that are obtained numerically are shown with solid lines.
Dispersion curves that are obtained using analytical expression are shown with dashed lines.
b,c) Spin-wave spectra of S/F/S based magnonic crystals that are formed in MSSW geometry with the lattice parameter $a=1$~$\mu$m.
Inserts in (b,c) show schematic illustrations of the considered magnonic crystals.
}
\label{muMag}
\end{center}
\end{figure*}

The effect of the superconducting proximity in S/F/S systems on magnetization dynamics can be effective in magnonics for variation of the FMR frequency or for modulation of the spin-wave velocity.
In this section, micromagnetic simulations are employed\cite{Handbook} for calculation of spin-wave spectra for S/F/S-based continuous films and periodic structures in the magnetostatic surface wave (MSSW) geometry\cite{Deorani_CAP_14_S129,Serga_JPDAP_43_264002}, following Refs.~\cite{Kim_JPDAP_43_264004,Dvornik_TAP_125_101,Venkat_IEEE_49_524}.
The following micromagnetic parameters of studied F-layers are considered, which correspond to S1 sample:
thickness of F-layer $d=20$~nm, the saturation magnetization $\mu_0 M_s=1$~T, the anisotropy field $H_a=0$, the applied field $\mu_0 H=0.02$~T, the
exchange stiffness constant $A = 1.3 \times 10^{-11}$~J/m, and the gyromagnetic ratio $\mu_0\gamma=2.21 \times 10^5$~m/A/s. 
The excitation field pulse has the maximum frequency $f_{max} = 30$~GHz, the gaussian spatial profile with the width at half-maximum of 200 nm, and the amplitude of 0.001Ms.
In simulations, the diamagnetic (Meissner) contribution of S-subsystem on magnetization dynamics was accounted via the method of images \cite{Golovchanskiy_AdvFunctMater_28_1802375,Golovchanskiy_JAP_124_233903} in case of continuous S-layers and via the diamagnetic representation of superconductors\cite{Golovchanskiy_AdvSci_6_1900435,Golovchanskiy_JAP_127_093903} in case of a finite-size S-elements.
The effect of the superconducting proximity in S/F/S is represented by a local uniaxial anisotropy field $\mu_0 H_s=0.07$~T that corresponds to S1 sample (see Fig.~\ref{Fit1}a).

Figure~\ref{muMag}a collects simulation results for continuous thin films.
Blue solid curves show a typical dispersion curve for MSSW in the plain F-film that is obtained with simulations in absence of any contribution from S-subsystem.
Simulation results are well confirmed by the analytical dispersion relation \cite{Golovchanskiy_JAP_124_233903}, shown with blue dashed curves.
Red solid and dashed lines show dispersion curve of MSSW in the S/F bilayer in presence of magnetostatic interaction between the S and the F subsystems.
The magnetostatic interaction is accounted using the method of images \cite{Golovchanskiy_AdvFunctMater_28_1802375,Golovchanskiy_JAP_124_233903}.
It shows that in presence of magnetostatic interaction the dispersion is nonreciprocal: the frequency pass-band for positive wavenumbers is approximately doubled as compared to the pass-band for negative wavenumbers.
The nonreciprocity is a known property of MSSWs, which emerges due to asymmetry of the ferromagnetic film across its thickness or due to asymmetry of its surrounding (see Ref.~\cite{Golovchanskiy_JAP_124_233903} for details).
Purple solid and dashed lines show dispersion curve of MSSW in the S/F/S three-layer in presence of the proximity-induced uniaxial anisotropy $H_s$ but absence of magnetostatic interaction between the S and the F subsystems.
It shows that at zero wavenumber the difference in frequencies between the plain film and the film with uniaxial anisotropy is maximum and corresponds to the difference in FMR frequencies.
Upon increasing the wavenumper the difference in frequencies reduces.
Black solid and dashed lines show dispersion curve of MSSW in the S/F/S three-layer in presence of both the proximity-induced uniaxial anisotropy $H_s$ and of the magnetostatic interaction between the S and the F subsystems.
Comparison of these curves with dispersions in plain F film, in S/F bilayer and in F film with proximity-induced uniaxial anisotropy $H_s$ indicates that both the proximity-induced anisotropy and the magnetostatic screening affect the kinetics of spin waves.
The magnetostatic screening is the dominating effect on spin-wave velocity at the range of higher wavenumbers, while the proximity-induced anisotropy is dominating in vicinity to 0 wavenumbers.

Figures~\ref{muMag}b,c show the spin-wave spectrum of S/F/S-based magnonic crystals where periodicity of the dispersion is reached by periodic location of S/F/S-three-layered areas.
Figure~\ref{muMag}b shows the spectrum of the hybrid magnonic crystal that consists of alternating F and S/F/S sections (see the inset) with the lattice period $a=1$~$\mu$m, the width of F-section 0.5~$\mu$m, and the thickness of S-layers 120~nm.
Calculating this spectrum both the diamagnetic representation of S-stripes\cite{Golovchanskiy_AdvSci_6_1900435,Golovchanskiy_JAP_127_093903} and local S/F/S-induced anisotropy are considered.
The spectrum can be characterized as conventional one: it consist of allowed and forbidden bands, the forbidden bands are opened at Brillouin wavenumbers $1/2a$.
The width of band gaps reduces at higher frequencies. 
For instance, the first (lower-frequency) band gap is of width about 1.8~GHz, and the second band gap is of width 1~GHz.

Figure~\ref{muMag}c shows the spectrum for an alternative realization of the hybrid magnonic crystal, which consists of alternating F/S and S/F/S sections (see the inset).
Lower S-subsystem forms a continuous layer, so the structure is spatially asymmetric in respect to the $z$-axis.
For this structure similar geometrical parameters are considered: the lattice period $a=1$~$\mu$m, the width of F/S-section 0.5~$\mu$m, and the thickness of the upper S-layers 120~nm.
Calculating this spectrum the diamagnetic representation of S-stripes\cite{Golovchanskiy_AdvSci_6_1900435,Golovchanskiy_JAP_127_093903}, the image method\cite{Golovchanskiy_AdvFunctMater_28_1802375,Golovchanskiy_JAP_124_233903} have been used for finite-size and continuous superconducting elements, respectively.
The effect of the proximity in S/F/S sections is represented by the same local anisotropy $H_s$.
The spectrum for this spatially asymmetric structure is different.
The spectrum consist of allowed and forbidden bands.
The forbidden bands are of similar width as in Fig.~\ref{muMag}b: the first (lower-frequency) band gap is of width about 1.7~GHz, and the second band gap is of width 0.9~GHz.
However, spatial asymmetry induces nonreciprocity of the spectrum and indirect location of band gaps away from Brillouin wavenumbers.

It should be noted that in both cases the effect of the proximity in S/F/S sections is dominating for formation of band gaps:
in absence of this effect forbidden bands are not obtained.
This can be explained by a rather weak diamagnetic response of S-subsystems on spin waves with considered wavelength.
However, diamagnetic response of S-subsystems does affect frequency and wavenumber position of allowed and forbidden bands .

As a final remark we should note that for magnonic crystals with thicker F-layers the bandwidth of the forbidden bands is expected to increase correlating with the zero-temperature anisotropy field.
In particular, the bandwidth of the forbidden bands for a S/F/S-based magnonic crystal with F-layer of thickness of a few hundreds of nm is expected to be comparable with values for bi-componental magnonic crystals\cite{Ma_APL_98_153107,Ma_JAP_111_064326}.

\section{Conclusion}

Summarizing, magnetization dynamics is studied in superconductor/ferromagnet/superconductor multilayers in presence of superconducting proximity.
It is shown that superconductivity in S/F/S three-layers shifts the FMR to higher frequencies.
Presence of both S-layers and proximity at both S/F interfaces are mandatory for the phenomenon.
The frequency shift is quantified by the proximity-induced positive in-plane anisotropy $H_a$ and by a drop of effective magnetization $M_{eff}$.
Both $H_a$ and the drop of $M_{eff}$ are comparable.
The phenomenon shows a dependence on the thickness of the F-layer: for thicker F-layer the shift of the FMR frequency is substantially stronger.
For two studied samples with thickness of the F-layer 45 and 350 nm the highest natural FMR frequencies and corresponding anisotropies are reached among in-plane magnetized ferromagnetic systems. 
At the current stage even a qualitative explanation of the effect of superconducting proximity in S/F/S systems on magnetization dynamics is unavailable.

Application of the proximity-induced anisotropies for manipulation with the spin-wave spectrum is demonstrated for continuous films and periodic magnonic crystals.
In general, presence of proximity-induced anisotropies in continuous films increase the phase velocity of spin waves especially at low wavenumbers.
In case of periodic structures, presence of alternating proximity-induced anisotropies ensure formation of forbidden bands for spin-wave propagation of width in GHz frequency range.

\section{Acknowledgments}

The authors acknowledge Prof.~V.~M.~Krasnov for fruitful discussions and for critical reading of the manuscript. 
This work was supported by the Ministry of Science and Higher Education of the Russian Federation, by the Russian Science Foundation, and by the Russian Foundation for Basic Research.



\bibliographystyle{apsrev}
\bibliography{A_Bib_SFS}

\end{document}